\documentclass[pre,twocolumn,notitlepage,longbibliography,amsmath,amssymb,floats,superscriptaddress,nofootinbib,10pt]{revtex4-1}

\pdfoutput=1

\AtBeginDocument{%
    \newwrite\bibnotes
    \def\bibnotesext{Notes.bib}
    \immediate\openout\bibnotes=\jobname\bibnotesext
    \immediate\write\bibnotes{@CONTROL{REVTEX41Control}}
    \immediate\write\bibnotes{@CONTROL{%
    apsrev41Control,author="08",editor="1",pages="1",title="0",year="1"}}
     \if@filesw
     \immediate\write\@auxout{\string\citation{apsrev41Control}}%
    \fi
}%

\usepackage{graphicx}
\usepackage{dcolumn}
\usepackage{bm}
\usepackage{revsymb}
\usepackage[usenames]{color}
\usepackage{subfigure}
\usepackage{color}
\usepackage{physics}
\usepackage{amsmath}
\usepackage{cancel}

\usepackage[usenames]{color}
\usepackage{amsfonts}
\usepackage{graphicx}
\usepackage{dcolumn}
\usepackage{bm}
\usepackage{revsymb}
\usepackage[usenames]{color}
\usepackage{subfigure}
\usepackage{color}
\usepackage{physics}
\usepackage{amsmath}
\usepackage{amssymb,bbm}
\usepackage{soul}

\newcommand{\p}{p}
\newcommand{\q}{q}
\newcommand{\kk}{k}
\newcommand{\ph}{\hat{p}}
\newcommand{\qh}{\hat{q}}
\newcommand{\kh}{\hat{k}}
\newcommand{\lh}{\hat{\ell}}
\newcommand{\lambdah}{\hat{\lambda}}
\newcommand{\Lh}{\hat{\Lambda}}

\usepackage{xcolor}
\newcommand{\Comment}[3]{\par\noindent\textcolor{#1}{\llap{\footnotesize #2 }\fbox{\parbox{0.98\linewidth}{\textsf{\footnotesize #3}}}}\par}

\newcommand{\RN}[1]{\Comment{bblue}{RL}{#1}}

\newcommand{\bea}{\begin{eqnarray}}
\newcommand{\eea}{\end{eqnarray}}

\newcommand{\shell} {\Lambda \rightarrow \Lambda / b}
\newcommand{\hatshell} {\hat{\Lambda} \rightarrow \hat{\Lambda} / b}

\definecolor{nblue}{RGB}{28,130,185}

\definecolor{cgreen}{RGB}{76,153,0}

\definecolor{myorange}{RGB}{245,156,74}

\usepackage{hyperref}
\hypersetup{
  colorlinks=true,
  citecolor=magenta,
  urlcolor=-myorange
}

\newcommand{\quotes}[1]{``#1''}

\usepackage{xcolor}
\definecolor{ogreen} {RGB}{71,191,145}
\definecolor{bblue} {RGB}{137,207,240}

\definecolor{edit} {RGB}{123,150,145}
\definecolor{purple} {RGB}{148,0,211}

\newcommand{\rn}[1]{{\color{bblue}{{#1}}}}

\begin{document}

\title{Fractonic Berezinskii–Kosterlitz–Thouless transition \\ from a renormalization group perspective }

\author{Kevin T. Grosvenor}
\email{grosvenor@lorentz.leidenuniv.nl}
\affiliation{Instituut-Lorentz, Universiteit Leiden, P.O. Box 9506, 2300 RA Leiden, The Netherlands.}

\author{Ruben Lier}
\email{rubenl@pks.mpg.de}
\affiliation{Max Planck Institute for the Physics of Complex Systems, 01187 Dresden, Germany}
\affiliation{W\"{u}rzburg-Dresden Cluster of Excellence ct.qmat, Germany}

\author{Piotr Sur\'{o}wka}
\email{piotr.surowka@pwr.edu.pl}
\affiliation{Institute for Theoretical Physics, Wroc\l{}aw  University  of  Science  and  Technology,  50-370  Wroc\l{}aw,  Poland}
\affiliation{Institute for Theoretical Physics, University of Amsterdam, 1090 GL Amsterdam, The Netherlands}
\affiliation{Dutch Institute for Emergent Phenomena (DIEP), University of Amsterdam, 1090 GL Amsterdam, The Netherlands}

\begin{abstract}
Proliferation of defects is a mechanism that allows for topological phase transitions. Such a phase transition is found in two dimensions for the XY-model, which lies in the Berezinskii-Kosterlitz-Thouless (BKT) universality class. The transition point can be found using renormalization group analysis. We apply renormalization group arguments to determine the nature of BKT transitions for the three-dimensional plaquette-dimer model, which is a model that exhibits fractonic mobility constraints. We show that an important part of this analysis demands a modified dimensional analysis that changes the interpretation of scaling dimensions upon coarse-graining. Using this modified dimensional analysis we compute the beta functions of the model and predict a finite critical value above which the fractonic phase melts, proliferating dipoles. Importantly, the transition point is found through a renormalization group analysis that accounts for the phenomenon of UV/IR mixing, characteristic of fractonic models.
\end{abstract}

\maketitle

\section{Introduction}

Quantum excitations with mobility constraints constitute new phases of matter called fractons. In the quest of understanding properties of these phases and their experimental realization it is necessary to develop their macroscopic description and identify unique properties that distinguish them from other phases of matter studied in the past. The framework that allows for a systematic investigation of physical systems at different scales is called the renormalization group (RG). Unfortunately a direct application of this framework to fracton models is faced with difficulties due to the so-called UV/IR mixing phenomenon \cite{uvirmix1,uvirmix2,you2020fracton,you2021fractonic}, which in essence means that, depending on the chosen direction, low-energy modes can have very high momenta. As a consequence it was suggested that renormalization group is not applicable to fracton phases \cite{Zhou:2021wsv,moessneryizhi}. However, it was recently argued that this difficulty can be circumvented by adapting the integration of the high-energy modes to the symmetries exhibited by the fracton models \cite{ethanlake}. 

 Renormalization group analysis of the XY model reveals that it lies in the universality class of the Berezinskii-Kosterlitz-Thouless (BKT) transition \cite{Kosterlitz_1973,berezinskii1971destruction}. A distinctive feature of this universality class is the critical temperature that governs the proliferation of free topological defects. Applying a similar analysis to fractonic theories can potentially provide a novel diagnostic of universality classes in theories with mobility constraints. Our goal is to refine the framework of renormalization group to study the proliferation of defects in a theory with mobility constraints. 
 
 A new universality class of the BKT type was recently anticipated in the context of superfluids and plaquette-dimer liquids \cite{moessneryizhi}. Dimer models represent lattice systems with degrees of freedom on the links instead of the nodes \cite{dimer1,dimer2,dimer3,dimer4,dimer5,dimer6,dimer7,dimer8,dimer9,dimer10}. Such models originate from the quest of understanding magnetic materials and are used to shed light on valence bond liquids or classical spin ice. In constructing these dimer models one imposes a constraint on the dimers, namely that each site form a dimer with only one of its neighbors. Sites that violate this condition are associated with defects. A site that is not attached to any dimer is called a monomer. A single monomer cannot move alone, while a pair of
monomers between links can only move along the transverse direction. A generalization of simple dimers involves trimers and plaquettes. Crucially, a class of plaquette-dimer models can be mapped to electrostatics with higher-rank tensor electric fields, considered by Pretko in the context of gapless fractons \cite{pretko_generalized_2017,pretko_subdimensional_2017}. In addition the low-energy effective theory is governed by the physics of defects, i.e., singular configurations of the fields \cite{seiberg_exotic_2020}. The goal of the present paper is to employ renormalization group analysis to study the transition from the liquid phase to the ordered configuration. Our approach can be applied to a variety of fracton models using the powerful technique of the renormalization group.


\section{Fractonic plaquette-dimer model} 
In this work we study a dual fractonic model obtained in Ref.~\cite{moessneryizhi} from the point of view of momentum shell RG. The model follows from Villain-dualizing a plaquette-dimer model with compact fields. We can formulate the Lagrangian density of this dual model as 
\begin{align} \label{partitionfunccc}
      \mathcal{L} &    =  \frac{ \kappa_{xy}}{2}    (   \hat{\Delta}_x \hat{\Delta}_y   h )^2    + \frac{ \kappa_{z}}{2}  (  \hat{\Delta}_z h )^2 \notag \\
      &\quad + 2  \sum_I  \alpha_I \cos(2 \pi f_I   ) ~~,
\end{align} 
where $\hat{\Delta}_{\mu}$ is a discrete derivative that acts on a general function $j (\mathbf{x}_j )$ as
\begin{align}
    \hat{\Delta}_{\mu} j(\mathbf{x}_j)  =   
a_{\mu}^{-1} \left(  j(\mathbf{x}_j  + a_{\mu} \mathbf{e}_{\mu} )  - j(\mathbf{x}_j   )  \right)  ~~ ,   
\end{align}
where $\mu \in (x,y,z)$, $a_{\mu}$ are the lattice constants, and $\mathbf{e}_{\mu}$ is a unit vector. The cosine terms in eq.~\eqref{partitionfunccc} contain a $h$-dependent functions $f_I$ and have a corresponding \quotes{fugacity} $\alpha_I$. For the fractonic-dimer plaquette model, the dual theory contains the functions
\begin{align} \label{feiuhe982}
    f_I = \{ h , a_x \hat \Delta_x h  , a_y \hat \Delta_y h   \} ~~ , ~~  I \in \{0,x,y  \} ~~, 
\end{align}
These $f_I$ are functions that enter through a cosine as they represent topological defects. Specifically, the $I$=0 term represents a monopole defect that can only move along the $z$-direction. Such a cosine term is also present for the Sine-Gordon model dual to the XY model \cite{kadanoff,herbut_2007}. The $I=x,y$ terms represent dipole defects that can only move in the direction orthogonal to the dipole direction ($x$ and $y$ directions, respectively) \cite{moessneryizhi}. 
The Lagrangian enters the partition function as
\begin{align} \label{eq:partitionfunc}
\mathcal{Z} = \int_{-\infty}^{\infty}  \prod_i   d h (\mathbf{x}_j)  \, \exp \bigl[ - \Omega_a  \sum_i \, \mathcal{L} (h (\mathbf{x}_j) ) \bigr] ~~  ,   
\end{align}
where the index $i$ runs through all the sites of the cubic lattice and $\Omega_a$ is the volume of the lattice site.\footnote{By putting $\Omega_a$ in the exponential we made sure that the Lagrangian of eq.~\eqref{partitionfunccc} is indeed a density, despite it being summed over a lattice. This will be of use in the proceeding RG analysis.} We study the system at static equilibrium, which is why the time integral does not appear in the action in eq.~\eqref{eq:partitionfunc}, having been canceled by the inverse temperature in Euclidean signature. 

Characteristic of fractonic models, the Gaussian term in eq.~\eqref{partitionfunccc} has a peculiar dispersion, which can be obtained by performing a Fourier transform that leads to $  \frac{\kappa_{xy}}{2}    ( \hat   \Delta_x  \hat \Delta_y   h )^2    + \frac{\kappa_{z}}{2}  ( \hat  \Delta_z h )^2
   \rightarrow    \frac{1}{2} 
 \epsilon_\mathbf{p} h^2$, with 
\begin{align}
 \epsilon_\mathbf{p}   &=    16  \kappa_{xy}   \sin^2 ( a_x p /2   ) \sin^2 (  a_y q /2 )    \notag \\
&\quad + 4 \kappa_{z}   \sin^2 ( a_z k /2  ) ~~,  
\label{eq:dispersion}
\end{align}
where we defined $\mathbf{p} = (\p,\q,\kk)^{\intercal}$. To simplify the dispersion, we assume that
\begin{align} \label{eq:simplification}
  a_x p  \ll 1  ~~ , ~~ a_y q  \ll 1  ~~   , ~~a_z k   \ll 1  ~~ , 
\end{align}
so that eq.~\eqref{eq:dispersion} turns into\footnote{This simplification cannot be generally valid for fractonic models, as UV/IR mixing makes it so that even at low energies eq.~\eqref{eq:simplification} can be violated. We show however in App.~\ref{sec:discrete} that assuming eq.~\eqref{eq:simplification} merely leads to a quantitative deviation.}
\begin{align} \label{eq:dispersionsimple}
  \epsilon_\mathbf{p}  \approx    \kappa (p^2 q^2 + k^2 )    ~~ ,  
\end{align}
where we defined\footnote{The definition of $\kappa$ in eq.~\eqref{eq:definition} is not without loss of generality as $\kappa_{xy}$ and $\kappa_{z}$ are independent. Eq.~\eqref{eq:definition} is introduced as it simplifies the RG picture. At a later stage the independent $ \kappa_{xy}$ and $ \kappa_z$ will be reintroduced when needed.}
\begin{align} \label{eq:definition}
 \kappa =   \kappa_{x y } =  \kappa_z   ~~ . 
\end{align}
Since the dispersion of eq.~\eqref{eq:dispersionsimple} vanishes when $\p =0$ or $\q =0$ one has to take note of short-wavelength effects even at low energies. Because of this is argued in Ref.~\cite{moessneryizhi} that this model is beyond the renormalization group paradigm. By considering the correlation functions, those authors were nevertheless able to derive a critical point where the low-energy theory is no longer described by a free fractonic theory but instead by a proliferation of dipoles. Because this proliferation of defects destroys the quasi-long-range order, it is reminiscent of the BKT transition \cite{Kosterlitz_1973,berezinskii1971destruction}. Corresponding to this phase transition is a critical $\kappa_{c}$, above which the coefficients $\alpha_x$ and $\alpha_y$ are relevant and grow large in the IR. When this happens, the field $h$ arranges itself to be in the valley of the cosines with $I = x,y $, so that at low energies the cosines can be expanded and one is left with a three-dimensional sine-Gordon model \cite{kadanoff} that does not have any fractonic properties. In this work we show that it is possible to do momentum shell RG for the model of eq.~\eqref{partitionfunccc} by using an RG procedure which is an extension of the RG procedure in Ref.~\cite{ethanlake}, where RG for the exciton Bose liquid \cite{balentsringexchange,cenkexu} is considered, which is a quantum model that also suffers from UV/IR mixing. With this approach the momentum shells are along the constant energy surface of the fractonic dispersion (see Fig. \ref{fig:esurfaces}). Because of this, one flows towards the axes as opposed to the origin, and one thus avoids issues related to UV/IR mixing. In the following section, we discuss this approach and specifically the nature of the dilatation operator for this RG procedure. We then use this to derive the critical point $\kappa_c$. With this approach we are furthermore enabled to compute the screening effect that the cosine terms have on $\kappa$, i.e., the inverse of the \quotes{fractonic spin-stiffness}, which we find to be absent.


\section{RG procedure}

The RG procedure of integrating out high-energy modes in a way which is adapted to the fractonic plaquette-dimer model was first proposed and implemented in \cite{ethanlake}. However, the crucial subsequent step of rescaling the low-energy modes back up in order to return to the original form of the action, but with renormalized parameters, was still done the usual way, which is \emph{not} adapted to the fractonic model. In this section, we describe the appropriate modification to this second step. In essence, what we are pointing out here is that the concept of \emph{dimensional analysis} itself is modified near the fractonic free-field fixed point as compared with the homogeneous one. This will obviously influence what we mean by \emph{relevant}, \emph{irrelevant}, and \emph{marginal} in the context of RG.

\begin{figure}[t]
    \centering
    \includegraphics[width=0.35\textwidth]{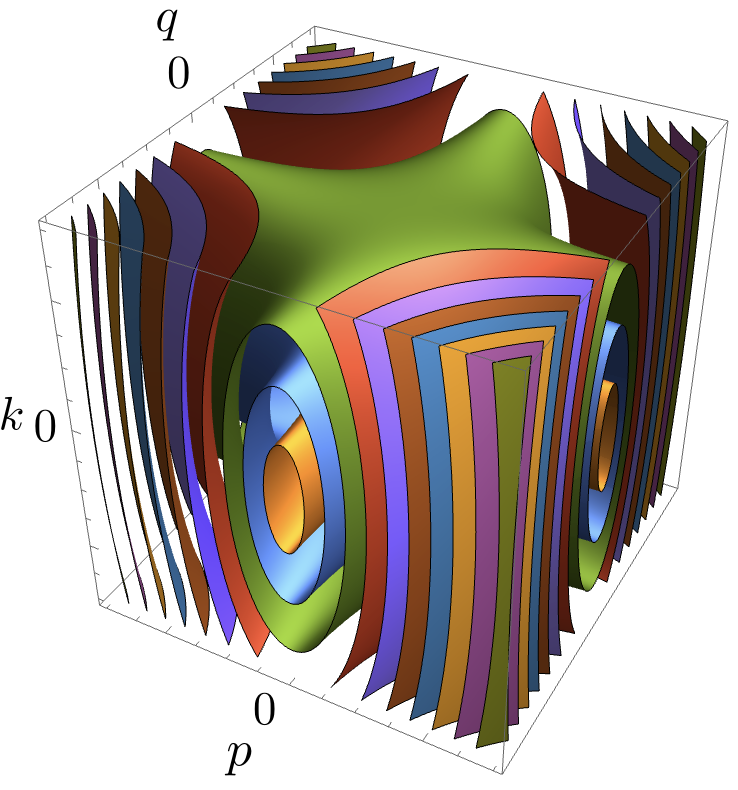}
    \caption{Constant-energy surfaces.}
    \label{fig:esurfaces}
\end{figure}

The most natural way to make manifest the concept in RG of ``flowing to the IR'' is to integrate out high-\emph{energy} shells. These contain only modes with energies between some high-energy cut-off $\Lambda$ to some slightly lower energy scale $\Lambda / b$, where $b$ is a number slightly greater than 1. We emphasize that this is a high-energy shell and not a high-momentum shell, as it is often called; it simply happens to be the case that high energy and high momentum are usually equivalent. This is \emph{not} the case for the fractonic phase. Nevertheless, the procedure is operationally the same: we integrate out modes with energy between $\Lambda$ and $\Lambda / b$. In momentum space, $(\p, \q, \kk )$, the constant-energy surface is given by $\sqrt{(\p \q)^2 + \kk^2} = \Lambda$ and depicted in Fig. \ref{fig:esurfaces}. Whereas the zero-energy locus is usually just one point, namely the origin in momentum space, in our case, it is the union of the $p$- and $q$-axes.

Normally, the constant-energy surfaces are concentric spheres and the RG flow is towards the origin. Therefore, the RG flow is described by the dilatation operator in momentum space $D = \mathbf{p} \cdot \nabla_{\mathbf{p}}$. Thus, after integrating out a shell between energies $\Lambda$ and $\Lambda / b$, we rescale momenta precisely by the dilatation operator in order to bring $\Lambda / b$ back up to $\Lambda$. In this regard, what is meant by the dimension of an operator or a parameter is really just its eigenvalue with respect to dilatation.

In our case, it no longer makes sense to define scaling dimension with respect to the dilatation operator $D$ since that does \emph{not} describe the RG flow towards the IR anymore. Instead, we define a modified dilatation operator $\tilde{D}$, which describes the flow from one constant-energy surface to another. First, define the modified momentum vector
\begin{equation} \label{feoie209209}
    \tilde{\mathbf{p}} = \bigl( [\p] \, \p , [\q] \, \q , [ \kk ] \kk \bigr)^{\intercal}~~,
\end{equation}
where $[\p], [\q], [\kk]$ denote the dimensions of the momentum components. The new dilatation operator is
\begin{equation}
    \tilde{D} = \tilde{\mathbf{p}} \cdot \nabla_{\mathbf{p}}~~,
\end{equation}
under which the eigenvalues of $\p$, $\q$ and $\kk$ are, by construction, their dimensions. We ask that $\tilde{D}$ be orthogonal to the constant-energy surfaces. Furthermore, we have the freedom to set the dimension of any one component to 1, which then fixes the remaining two. It is natural to set $[\kk] = 1$, which fixes the dimensions to be
\begin{align} \label{eiewh209092}
    [ \p ] &= \frac{\q^2}{\p^2 + \q^2}~~, &%
    [ \q ] &= \frac{\p^2}{\p^2 + \q^2}~~, &%
    [ \kk ] &= 1~~. 
\end{align}
This means that if we rescale the energy $\Lambda \rightarrow b \Lambda$, then the momenta are rescaled according to $\p \rightarrow b^{[\p]} \p$, and similarly for $\q$ and $\kk$. This is quite different and exotic compared with the standard dimensional analysis, which would say that $\p$ and $\q$ both have dimension $\frac{1}{2}$. Of course, in either case, the dimension of the product $pq$ is still given by
\begin{align}
    [ \p \q ] =1 .  \label{scalingproducttt}
\end{align}

Since the meaning of classically relevant, irrelevant, and marginal depends on dimensional analysis, the above modified definition of dimensions has some rather important consequences. One issue is the question of whether gradient operators without fractonic properties, such as $(\partial_x h)^2$ and $( \partial_y h )^2$, will get generated. The naive form of dimensional analysis would conclude that these operators are relevant compared with the operator $( \partial_x \partial_y h)^2$. However, with respect to the modified dimensional analysis appropriate near the fractonic fixed point, $(\partial_x h)^2$ and $( \partial_y h )^2$ do not even have constant dimensions and, in fact, their highest dimension is 2, which is the same as the dimension of the operator $( \partial_x \partial_y h )^2$. This may seem to suggest that these operators are no more relevant than the fractonic one. However, consider, for example, the dimension of the operator $( \partial_y h )^2$ in the region near the $q$-axis, which is where it could destroy UV/IR mixing. Here, this dimension vanishes and from the point of view of the potential destruction of UV/IR mixing, this operator is even more relevant than in the naive dimensional analysis. We conclude that the fractonic theory is highly susceptible to being destroyed by simple gradient terms without fractonic properties, which is why it is so important that the UV theory is pristine. Because of this, dimensional analysis alone cannot determine whether or not the ordinary gradient terms do in fact get generated. We then ask what kind of theory the cosine perturbations induce in the IR. To answer this question, we now perform the RG analysis along the lines we described above.


\section{Fractonic plaquette-dimer melting by the proliferation of dipoles}

With the RG procedure just described, we can now derive the critical point where the fractonic dimer-plaquette liquid becomes unstable, which was discussed in Ref.~\cite{moessneryizhi}. This critical point is due to the cosine operators in eq.~\eqref{partitionfunccc} with $I=x$ and $I=y$ becoming relevant. In the appendices, following Refs.~\cite{Herbut_2003,seradjeh,herbut_2007}, we consider the effect of a general cosine term on the renormalization of the coefficients when one splits the field $h$ into
\begin{align}
     h   = h^-  + h^{+}  ~~ ,  \label{divide}
\end{align}
where $h^{+}$ corresponds to the high-energy modes in the momentum shell that are being integrated out, leading to a renormalization of the coefficients for the low-energy theory with modes $h^-$. Taking $I=x$, we show in App. \ref{partone} that the fugacity coefficient $\alpha_x$ experiences the following renormalization:
\begin{align} 
     \alpha_x   (b)  = b^2    \alpha_x   e^{- \frac{1}{2}  g^{+}_{(xx)} ( 0 )}  ~~ ,   \label{wickforfuga}
\end{align}
where $g_{(IJ)}^{+} ( 0 )$ is the correlator of the high-energy modes for operators $f_I $ and $f_J$. The factor $b^2$ is consistent with the rescaling for the momenta previously described, specifically eq.~\eqref{eiewh209092}, which determines the dimension of $\alpha_x$ to be 2. We thus must compute\footnote{In eq.~\eqref{feihew039398398}, we have made the same assumption as in eq.~\eqref{eq:simplification}, which enables us to linearize the discrete derivatives both in the kinetic terms as well as in the $\cos ( 2 \pi f_I )$ terms. The case without this assumption is considered in App.~\ref{sec:discrete}.}
\begin{align} \label{feihew039398398}
      g_{(xx)}^{+} ( 0 )  &=  \frac{a_x^2 }{\kappa}  \int_{\shell} \frac{ d^3   \mathbf{p} }{2 \pi }  \frac{\p^2 }{ ( \p \q  )^2 + \kk^2 }   ~~ . 
\end{align}
Let us first pass over to dimensionless variables 
\begin{align} \label{eq:undim}
    \ph &= \frac{a_x}{\pi} \, p~~, &%
    \qh &= \frac{a_y}{\pi} \, q~~, &%
    \kh &= \frac{a_x a_y}{\pi^2} \, k~~,
\end{align}
so that $| \ph |, | \qh | \leq 1$. Similarly, define the dimensionless cut-off,
\begin{equation}
    \Lh = \frac{\Lambda a_x a_y}{\pi^2}.
\end{equation}
With these definitions, eq. \eqref{feihew039398398} becomes
\begin{align} \label{gxxH0}
      g_{(xx)}^{+} ( 0 )  &=  \frac{\pi}{2 \kappa}  \int_{\hatshell} d^3   \hat{\mathbf{p}}
      \frac{\ph^2 }{ ( \ph \qh  )^2 + \kh^2 }   ~~ . 
\end{align}
We change variables from $( \ph , \qh , \kh )$ to $( \ph , \lh ,  \kh  )$, where $ \lh = \ph \qh$. The Jacobian for this change of variables is $\frac{1}{| \ph |}$:
%
%
%
\begin{align} \label{gxxH0b}
      g_{(xx)}^{+} ( 0 )  &=  \frac{4 \pi}{\kappa}  \int_{\substack{\hatshell \\ \kh , \lh , \ph > 0}} d \kh \, d \lh \, d \ph \,
      \frac{\ph}{\kh^2 + \lh^2 }   ~~ , 
\end{align}
where we used that by parity symmetry, we can restrict the integral to the positive octant and multiply by 8. Since $\lh = \ph \qh$ and the maximum value for $\qh$ is 1, the lower limit for $\ph$ is $\lh$. And, just as for $\qh$, the upper limit for $\ph$ is 1. Thus,
\begin{align} \label{gxxH0c}
      g_{(xx)}^{+} ( 0 )  &=  \frac{4 \pi}{\kappa}  \int_{\substack{\hatshell \\ \kh , \lh > 0}} \frac{d \kh \, d \lh}{\kh^2 + \lh^2} \int_{\lh}^{1} d \ph \,
      \ph \notag \\
      &= \frac{2 \pi}{\kappa}  \int_{\substack{\hatshell \\ \kh , \lh > 0}} \frac{d \kh \, d \lh}{\kh^2 + \lh^2} \, (1 - \lh^2 ) ~~. 
\end{align}
By $\kh \leftrightarrow \lh$ symmetry, we may replace the factor $1 - \lh^2$ in the integrand with $1 - \frac{\kh^2 + \lh^2}{2}$:
\begin{align} \label{gxxH0d}
      g_{(xx)}^{+} ( 0 )  &= \frac{2 \pi}{\kappa}  \int_{\substack{\hatshell \\ \kh , \lh > 0}} d \kh \, d \lh \biggl( \frac{1}{\kh^2 + \lh^2} - \frac{1}{2} \biggr) ~~. 
\end{align}
Now, convert to polar coordinates by setting
\begin{align}
    \kh &= \lambdah \cos \phi ~~, &%
    \lh &= \lambdah \sin \phi~~.
\end{align}
Since we have restricted to the positive quadrant in $\kh$ and $\lh$, the polar angle $\phi$ goes from 0 to $\frac{\pi}{2}$. Meanwhile, $\lambdah$ goes from $\Lh / b$ to $\Lh$. Therefore,
\begin{align} \label{gxxH0e}
      g_{(xx)}^{+} ( 0 )  &= \frac{2 \pi}{\kappa} \int_{0}^{\frac{\pi}{2}} d \phi \int_{\Lh /b}^{\Lh} d \lambdah \biggl( \frac{1}{\lambdah} - \frac{\lambdah}{2} \biggr) \notag \\
      &= \frac{\pi^2}{\kappa} \biggl[ \ln (b) - \frac{\Lh^2}{4} \biggl( 1 - \frac{1}{b^2} \biggr) \biggr]~~.
\end{align}
Plugging this into eq. \eqref{wickforfuga}, taking a derivative with respect to $\log (b)$, and setting $b = 1$ gives the RG flow:
\begin{align} \label{eq:exactbetaalphax}
    \frac{d \alpha_x (b)}{d \ln (b)} \biggr|_{b=1} = \biggl[ 2 - \frac{\pi^2}{2 \kappa} \, \biggl( 1 - \frac{\Lh^2}{2} \biggr) \biggr] \alpha_x~~.
\end{align}
As we have argued previously by our modified dimensional analysis, the UV/IR mixing is most in danger of being destroyed near the $p$-$q$ axes. To probe this region and to ensure that the momentum shell covers as wide an angular range as possible, we take the limit $\Lh \ll 1$. Thus, our final RG flow equation for $\alpha_x$ reads
\begin{align} \label{eq:limitbetaalphax}
    \frac{d \alpha_x (b)}{d \ln (b)} \biggr|_{b=1} \approx \biggl( 2 - \frac{\pi^2}{2 \kappa} \biggr) \alpha_x~~.
\end{align}
The RG flow equation for $\alpha_y$ is exactly the same, but with $\alpha_x \rightarrow \alpha_y$.

Hence, there is a critical value
\begin{equation} \label{eq:kappac}
     \kappa_c = \frac{\pi^2}{4}~~,
\end{equation}
at which $\alpha_x , \alpha_y$ are marginal at this order, below which $\alpha_x, \alpha_y$ are irrelevant, and above which $\alpha_x, \alpha_y$ become relevant. In other words, this operator \emph{does} destabilize the fractonic phase when $\kappa > \kappa_c$. This is precisely the process of dipole proliferation. When we compute higher-order screening effects on $\kappa$, we will see that this RG flow picture gets slightly more complicated, but the essential point remains that there is a region in parameter space where the fractonic phase is stable (see Fig. \ref{fig:rgflow}).

\begin{figure}[t]
    \centering
    \raisebox{-0.5\height}{\includegraphics[width=0.2\textwidth]{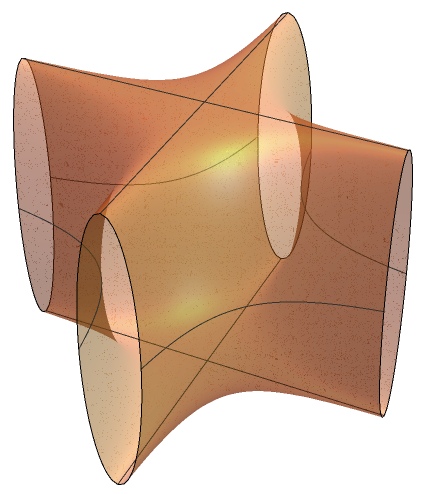}} \raisebox{-0.5\height}{\includegraphics[width=0.04\textwidth]{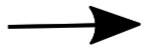}} \raisebox{-0.5\height}{\includegraphics[width=0.2\textwidth]{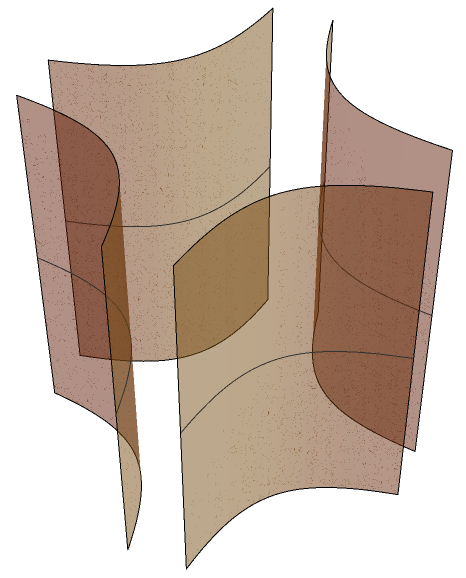}}
    \caption{The deformation of the momentum shell to one that is parallel to the $k$-direction.}
    \label{fig:surfacedeform}
\end{figure}

We will discuss the $k$-integral in more detail. In fact, the natural scale of $k$ as far as the shell is concerned is $\Lambda$ since a constant energy surface extends from $- \Lambda$ to $+ \Lambda$ in the $k$ direction. Therefore, in natural units, the width of the end-caps of the constant energy surface (i.e., where the surface hits the bounds set by the lattice constants) is of order $\Lh$, whereas the height is of order 1. In other words, locally near $k=0$, the surface is very narrow and almost vertical. This suggests the following simplification: deform the surface to be its intersection with the $k=0$ plane simply translated from $k = - \pi /a_z $ to $k = + \pi / a_z$, as shown in Fig. \ref{fig:surfacedeform}. Bear in mind that the elliptical holes on the original surface and the gaps in the deformed surface should, in truth, be very narrow. Our expectation is that the difference between performing the integral over the exact integration region and the deformed one should vanish in the limit $\Lh \ll 1$. Because of this deformation, the boundary conditions of the $k$-integral are now decoupled from the $p$- and $q$-integral, which are the momenta that are related to UV/IR-mixing. Because of this decoupling, there is no longer any reason not to work in the continuum limit for the $z$-direction, i.e. we take $a_z \rightarrow 0$, still keeping the $x$- and $y$-directions discrete. The integral simplifies to
\begin{align} \label{eq:integralsimplification1}
      \int_{\shell} d^3  \mathbf{p} \rightarrow \int_{\shell}   d p \, d q  \, 
 \int^{\infty }_{-\infty } d k     ~~ , 
\end{align}
and similarly for the hatted variables. Making this approximation gives
\begin{align} \label{eq:gHxx0f}
    g^{+}_{(xx)} (0) &= \frac{\pi}{2 \kappa} \int_{\hatshell} d \ph \, d \qh \int_{- \infty}^{\infty} d \kh \, \frac{\ph^2}{( \ph \qh )^2 + \kh^2} \notag \\
    &= \frac{2 \pi^2}{\kappa} \int_{\Lh}^{1} d \ph \, \ph \int_{\Lh / b \ph}^{\Lh / \ph} \frac{d \qh}{\qh} \notag \\
    &= \frac{\pi^2}{\kappa} ( 1 - \Lh^2 ) \ln (b) ~~.
\end{align}

\begin{figure}[t]
    \centering
    \includegraphics[width=0.45\textwidth]{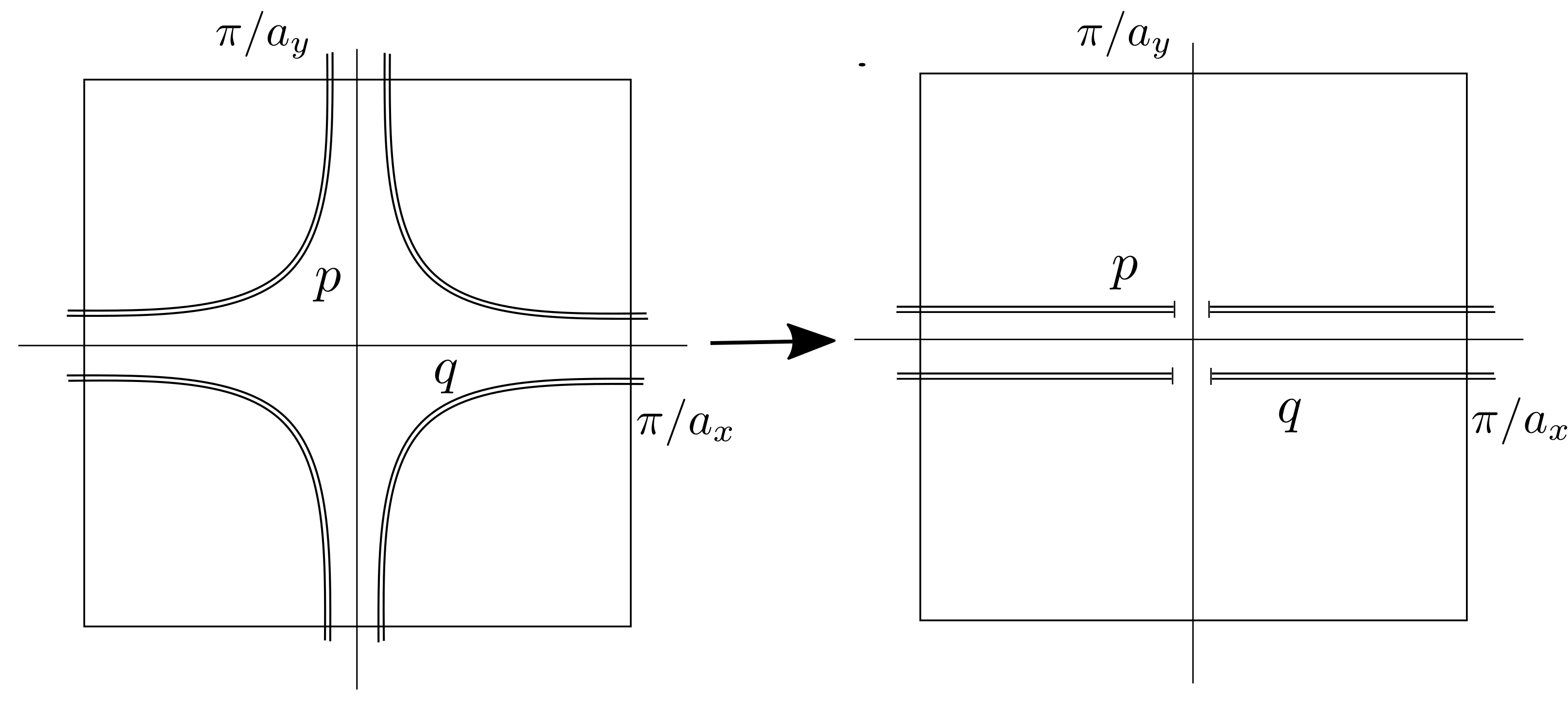}
    \caption{The deformation of the momentum shell corresponding to eq.~\eqref{eq:gHxx0f} but where the limits of the $\qh$ integral are \emph{not} divided by $\ph$.}
    \label{fig:esurfaces1111}
\end{figure}

Of course, this does not quite agree with eq. \eqref{gxxH0e}. However, as expected, the difference is of order $\Lh^2$ and vanishes in the $\Lh \ll 1$ limit.  
Note that due to the logarithmic nature of the $\qh$ integral, it makes no difference whether or not the limits of integration of $\qh$ are divided by $\ph$. Therefore, another simplification we can make is to flatten out the shell to be parallel to the $p$-axis, as depicted in Fig. \ref{fig:esurfaces1111}. The width of the gap near small $\ph$ is of order $\Lh$ and so the $\Lh \ll 1$ limit is equivalent to closing this gap. This fact is not so crucial here, but it will have very important consequences later on when we compute higher-order screening effects. In fact, it will play a critical role in our understanding of how to properly take the continuum limit of fractonic theories.

Finally, before we compute the RG flow of other parameters, we revisit the assumption eq.~\eqref{eq:simplification}, which allowed us to obtain the more workable dispersion of eq.~\eqref{eq:dispersionsimple}. This assumption works only when the lattice constant can be assumed to always be unimportant for the RG computations. However, due to the UV/IR mixing, the IR theory contains momentum modes near the axes that reach the momentum cutoff set by the inverse of the lattice constant, so that this assumption is never valid. Because of this, incorporating lattice effects modifies eq.~\eqref{eq:kappac} so that a different $\kappa^{(L)}_c$ is found. Specifically, the result found in App. \ref{sec:discrete} is given by
  \begin{align}
\kappa^{(L)}_c =2  ~~ ,    
  \end{align}
so that the lattice effects reduce the critical parameter by approximately $20\%$ relative to the continuous limit. 

\section{Monopole Effects}

The fugacity for the monopole operator is $\alpha_0$. We will now show that $\alpha_0$ can always be assumed to be irrelevant. The flow of $\alpha_0$ is given by
  \begin{align}
  \alpha_0   (b)  = b^2    \alpha_0   e^{- \frac{1}{2}  g^{+}_{(00)} ( 0 )}  ~~ .  \label{wickfoga}
  \end{align}
The quantity $g^{+}_{(00)} ( 0 )$ can be computed exactly in precisely the same manner as $g^{+}_{(xx)} (0)$. The crucial difference is that the $p^2$ term in the numerator of the integrand is no longer present. Where the integrand in the previous calculation vanished near small $p$, it now diverges like $\frac{1}{p}$ and thus contributes a logarithmic divergence. As computed in App.~\ref{sec:cos}, the exact result for $g^{+}_{(00)} (0)$ is
\begin{align}
    g^{+}_{(00)} (0) &= \frac{1}{\kappa} \int \frac{d^3 \mathbf{p}}{2 \pi} \, \frac{1}{(pq)^2 + k^2} \notag \\
&= \frac{2}{\kappa} \biggl[ \ln \biggl( \frac{2}{\Lh} \biggr) + \frac{1}{2} \ln (b) \biggr] \ln (b)~~.
\end{align}
The resulting RG flow reads
 \begin{align} \label{f3h030101-039889}
      \frac{\partial \alpha_{0} (b)}{\partial \ln (b)} \biggr|_{b=1} = \left[  2 - \frac{1}{\kappa} \ln \left( \frac{2}{\Lh} \right) \right] \alpha_0    ~~ .
  \end{align}
Because $\tilde \Lambda  \ll 1 $ we find that the coefficient $\alpha_0$ can always be considered to be irrelevant, confirming the result in Ref.~\cite{moessneryizhi}. To summarize, we have derived a similar prediction as that of Ref.~\cite{moessneryizhi}, namely that there is a finite critical value $\kappa_c$ above which the fractonic phase melts via dipole proliferation.We find only a difference in the precise value of $\kappa_c$. Crucially, we have established this result now firmly within the formalism of the renormalization group by working with momentum shells that adhere to UV/IR mixing.

Now that we know that the $I=0$ cosine term is irrelevant, one may still worry that while this irrelevant term flows to the IR shrinking rapidly, it acts as a dangerously irrelevant operator by giving rise to a new term in the Gaussian part in eq.~\eqref{partitionfunccc}. Let us imagine that this Gaussian part $\mathcal{L}_G$ gets modified so that it now includes the following $\gamma$-term: 
\begin{align} \label{eoiewo1111i}
\begin{split}
 \mathcal{L}^{\prime }_G &   =  \mathcal{L}_G + \frac{ \gamma}{2} \left[   (    \hat{\Delta}_x   h )^2 + (    \hat{\Delta}_y   h )^2   \right] ~~.
\end{split}
\end{align}
Even for small $\gamma$, this term violates the fractonic properties that were there for $\gamma=0$  and thus undoes the arguments related to the stability of the fractonic plaquette-dimer liquid as these arguments relied on UV/IR mixing. The flow equation for $\gamma$ is given in eq. \eqref{eq:gammaeq}: 
\begin{align} \label{eq:gamma}  
  & \gamma (b)
  = (2 \pi)^2   \alpha_0  ^2 (b)   \Omega_a   \sum_i    \,  (x_i^2+y_i^2 )     \Bigl( e^{g_{(00)}^+(\mathbf{x}_i )} - 1 \Bigr) \notag \\ 
  & \approx 2 \pi   \frac{\alpha_0  ^2 (b)}{\kappa }        \Omega_a   \sum_i     \int_{\shell}   d^3  \mathbf{p}   \frac{(x_i^2+y_i^2 )  e^{i \mathbf{p} \cdot  \mathbf{x}_i}}{ ( \p \q  )^2 + \kk^2 } ~~ .
\end{align}
This integral is complicated due to the nature of the cut-off surfaces. Therefore, we deform the shell as we did earlier: we integrate $k$ from $- \pi/a_z$ to $+ \pi / a_z$ and take the continuum limit for the $z$-direction, so that integrating the $z$-dimension out yields
\begin{align} \label{ee0920921}
 \gamma (b) &= \frac{  (2 \pi)^2  \alpha_0  ^2 (b)  }{\kappa } a_x a_y \notag \\
&\quad \times \sum_{i \in A_{xy}}  \int_{\shell}   d p \, d q  \,  \frac{(x_{i}^{2} + y_{i}^{2} )   e^{i p x_i + i q  y_i  }}{ ( \p \q  )^2  }~~ ,
\end{align}
where $A_{xy}$ runs through the spatial lattice on the $xy$-plane. The lattice constants $a_x$ and $a_y$ enter the integrand of eq.~\eqref{ee0920921} only to prevent the shell from touching the axes, and we can simply take the continuum limit so that the shells are asymptotic to the axes. This yields
\begin{align} \label{ee09209211}
\begin{split}
 &   \gamma (b) \\ & 
 =  \frac{  (2 \pi)^2  \alpha_0  ^2 (b)  }{\kappa }        \int  d x \, d y   \int_{\shell}   d p \, d q  \,  \frac{(x^2 + y^2 )   e^{i p x + i q  y  }}{ ( \p \q  )^2  }~~ .  \end{split}
\end{align}
We move to polar coordinates both for real space as well as for momentum space and first integrate over the angle $\theta$ in real space:
\begin{align} \label{ee0920jihiu92}
\begin{split}
 &   \gamma (b) \\  & 
 = \frac{  (2 \pi)^3  \alpha_0  ^2 (b)  }{\kappa }        \int_0^{\infty}  d r   \int_{\shell}   d \rho d \phi    \frac{r^3  J_0( r  \rho   )}{ \frac{1}{4}  \sin^2 (2 \phi )  \rho^3  }~~ .  \end{split}
\end{align}
We then change the integration variables as $\tilde \rho  =  \sqrt{\frac{1}{2}  | \sin (2 \phi ) | } \, \rho  $ and $\tilde r   =    r /  \sqrt{\frac{1}{2} | \sin (2 \phi ) |}$. Because we consider the continuum limit, the integral over $\phi$ is simply $2 \pi$:
\begin{align} \label{e328299289892}
  \gamma (b) &=  \frac{(2 \pi)^4  \alpha_0  ^2 (b)}{\kappa}        \int_0^{\infty}  d \tilde  r      \int^{\sqrt{\Lambda} }_{ \sqrt{\Lambda/b} }   d \tilde \rho \,  \frac{\tilde r^3  J_0( \tilde r \tilde  \rho   )}{   \tilde \rho^3  } ~~.
\end{align} 
The integral in eq.~\eqref{e328299289892} is divergent, but can be regularized by smoothing the momentum shell as follows:
\begin{align} \label{feiuheiu2090920921}
     \int^{\sqrt{\Lambda} }_{\sqrt{\Lambda/b} }  d \tilde \rho   \rightarrow   \int^{\infty}_{0}  d \tilde \rho   \left( \frac{\Lambda^{n} }{ \tilde \rho^{2n} + \Lambda^{n} } - \frac{(\Lambda/b)^{n} }{ \tilde \rho^{2n} + (\Lambda/b)^{n} }   \right) ~~,  
\end{align}
where for theories with quadratic correlators one can take $n=1 $ \cite{Herbut_2003,herbut_2007,seradjeh}. This turns a massless correlator for a momentum shell into the difference of massive correlators with effective masses $\sqrt{\Lambda}$ and $\sqrt{\Lambda/b}$ over the full momentum space, thereby avoiding an IR divergence. Similarly, the value for $n$ that regularizes eq.~\eqref{e328299289892} is $n=2$ which yields
\begin{align}
 \gamma (b) = 0 + \mathcal{O} \left(\log^2 (b)\right) ~~ .  \label{fehueiuh001010}
\end{align}
Therefore, the simple cosine operator is not in fact dangerous in that it does not generate terms which destroy the fractonic UV properties of the model. Note that the vanishing of $\gamma (b)$ relies only on the quartic nature of the correlation function that remains after the third spatial and momentum dimensions are integrated out in eq.~\eqref{ee0920921}. This means that for two-dimensional theories with quartic dispersion, such as the vector sine-Gordon model \cite{Zhai_2019}, which describes dislocation-induced melting of two-dimensional crystals \cite{Kosterlitz_1973,kthny1,kthny2,kthny3}, a similar argument precludes dangerously irrelevant cosine operators. The vanishing of the correction to the quadratic derivative term in the vector sine-Gordon model can also be derived using dimensional regularization in the same spirit as in the two-dimensional sine-Gordon model \cite{yanagisawa}.

\section{Higher-order screening effects} 
Now we consider the renormalization of quartic operators. For this, we first need to undo the simplification performed in eq.~\eqref{eq:definition} by splitting the $\kappa$ term up into
\begin{align}
   \frac{\kappa}{2} \Bigl(  (   \hat{\Delta}_x \hat{\Delta}_y   h )^2    &+   (  \hat{\Delta}_z h )^2 \Bigr)  \notag \\
& \rightarrow   \frac{\kappa_{xy}}{2}    (   \hat{\Delta}_x \hat{\Delta}_y   h )^2    + \frac{\kappa_{z}}{2}  (  \hat{\Delta}_z h )^2 ~~ .
\end{align}
Only $ \kappa_{xy} $ can receive corrections from screening effects related to the $I = x,y$ cosine terms. The flow equation for $\kappa_{xy}$ is given in eq. \eqref{eq:kappaxyeq}:  
\begin{align} \label{ee09209209huiiuh2}
  & \kappa_{xy} (b) - \kappa_{xy}= \delta \kappa_{xy} (b)  \notag \\ 
  &= 
   2 ( 2 \pi )^2  a_{x}^{2} \alpha_{x}^{2} (b) 
 \Omega_a \sum_i \, y_i^2 \Bigl( e^{g^{+}_{(xx)} (\mathbf{x}_i)} -1 \Bigr) + x \leftrightarrow y \notag \\
  &\approx \frac{4 \pi a_{x}^{4} \alpha_{x}^{2} (b)}{\kappa_{xy}}   \Omega_a \sum_i \, y_i^2 \int_{\shell} d^3  \mathbf{p} \, \frac{p^2   e^{i \mathbf{p} \cdot  \mathbf{x}_i}}{( \p \q  )^2 + \zeta \kk^2} \notag \\
&\quad + p,x \leftrightarrow q,y ~~,
  \end{align}
  with $\zeta = \kappa_{z} /  \kappa_{xy}$. The integral in eq.~\eqref{ee09209209huiiuh2} is divergent, and to regularize this integral is difficult, as the shell has a complicated structure. To make progress, we perform the same simplification as in Fig. \ref{fig:surfacedeform}, which is to deform the shell and take $a_z \rightarrow  0 $ so that we can integrate out $z$ and $k$. The result is \footnote{In this calculation, a factor of $p^2$ canceled between the numerator and the denominator after $z$ and $k$ were integrated out. Had we not made the assumption in eq.~\eqref{eq:simplification}, the factor that would have canceled between the numerator and denominator would have been $\frac{4}{a^2_x} \sin^2 (\frac{1}{2} a_x p )$. Either way, the result is the same. Therefore, whereas taking the more complicated and correct dispersion relation into account makes a quantitative difference for the flow of $\alpha_I$ with $I = x$ or $y$, as shown in App.~\ref{sec:discrete}, it does not make a difference for $\kappa_{xy}$.} 
\begin{align} \label{eq:kappaafterzkout}
    & \delta \kappa_{xy} (b) \notag \\
&= \frac{2  (2 \pi )^2 a_{x}^{5 } a_y  \alpha_{x}^{2} (b)}{\kappa_{xy}}   \sum_{i \in A_{xy}}  \,  y_i^2 \int_{\shell} d p  dq \, \frac{e^{i ( p x_i +  q y_i ) }}{q^2} \notag \\
&\quad + p,x \leftrightarrow q,y~~ .
\end{align}
where we see that $\zeta$ drops out completely. We again simplify the integrand in eq.~\eqref{eq:kappaafterzkout} according to Fig.~\ref{fig:esurfaces1111}, and introduce the dimensionless variables 
\begin{align}
    \tilde{q} &= \frac{\pi}{\Lambda a_x} \, q~~, &%
    \tilde{y} &= \frac{\Lambda a_x}{\pi} \, y~~,
\end{align}
to end up with
\begin{align} \label{eq:kappaafterzkoutee}
     \delta \kappa_{xy} (b)
    &= \frac{  (2 \pi)^5  a_x   a_{y} \alpha_{x}^{2} (b)}{2 \Lambda^3 \kappa_{xy}}    \Theta   \sum_{i \in L_y } \,  \tilde y_i^2 \int^{1}_{1/b }   d \tilde q \, \frac{e^{i   \tilde q  \tilde y_i  }}{\tilde q^2} \notag \\
&\quad + p,x \leftrightarrow q,y~~ , 
\end{align}
with 
\begin{align} \label{eq:theta}
  \Theta =  a_x   \sum_{i \in L_x }  \left\{ \int_{- \pi / a_x }^{- \Lambda a_y / \pi  } d p  + \int^{ \pi / a_x }_{\Lambda a_y / \pi  } d p  \right\}   e^{i p x_i }  ~~ . 
\end{align}
Now we consider the integral over the momentum shell in eq.~\eqref{eq:kappaafterzkoutee}. Because the integral over $q$ has no dependence on the lattice constant, we can now take the continuum limit. Note that the gap in the $p$ integral between $- \frac{\Lambda a_y}{\pi}$ and $+ \frac{\Lambda a_y}{\pi}$ closes in this limit for the $x$- and $y$-directions, i.e. we take $a_x =a_y=0$. Taking the continuum limit for the $x$- and $y$-directions turns the sums in eq.~\eqref{eq:kappaafterzkoutee} into integrals and closes the gap in the $p$-integral to give $\Theta = 2 \pi $. We thus find that eq.~\eqref{eq:kappaafterzkoutee} turns into
\begin{align} \label{eq:kappaafterzkoe}
     \delta \kappa_{xy} (b)  = &  \frac{ ( 2 \pi)^7   \alpha_{x}^{2} (b)}{4  
\Lambda^4  \kappa_{xy}}    \int d \tilde  y  \,  \tilde y^2 \int^{1}_{1/b }   d \tilde q \, \frac{e^{i   \tilde q  \tilde y_i  }}{\tilde q^2} \notag \\
& +  p, x \leftrightarrow  q,  y~~ , 
\end{align}
It is clear at this point that the only effect of the switch $p,x \leftrightarrow q,y$ is to change $\alpha_x$ to $\alpha_y$. Therefore, defining
\begin{align}
    \hat{\alpha}_x &= \frac{\alpha_x}{\Lambda^2}~~, &%
    \hat{\alpha}_y &= \frac{\alpha_y}{\Lambda^2}~~,
\end{align}
we can write
\begin{align} \label{eq:beforefinal}
    \delta \kappa_{xy} (b) &= \frac{(2 \pi )^7 \bigl[ \hat{\alpha}_{x}^{2} (b) + \hat{\alpha}_{y}^{2} (b) \bigr]}{4 \kappa_{xy}} \int d \tilde{y} \, \tilde{y}^2 \int_{\frac{1}{b}}^{1} d \tilde{q} \, \frac{e^{i \tilde{q} \tilde{y}}}{\tilde{q}^2}~~.
\end{align}
%
Eq.~\eqref{eq:beforefinal} can be regularized similarly to eq.~\eqref{feiuheiu2090920921}, by taking
\begin{align}
    \int_{\frac{1}{b}}^{1} d \tilde{q} &\rightarrow \int_{0}^{\infty} d \tilde{q} \biggl( \frac{\frac{1}{b^2}}{\tilde{q}^2 + \frac{1}{b^2}} - \frac{1}{\tilde{q}^2 + 1} \biggr)  ~~ ,
    \end{align}
so that the final result is
\begin{align} \label{eq:finalreslt}
    &\delta \kappa_{xy} (b) 
    = \frac{(2 \pi )^8 \bigl[ \hat{\alpha}_{x}^{2} (b) + \hat{\alpha}_{y}^{2} (b) \bigr]}{4 \kappa_{xy}} \biggl( 1 - \frac{1}{b^2} \biggr)~~.
\end{align}
The RG flow equation thus reads
\begin{align}
    \frac{\partial \kappa_{xy} (b)}{\partial \ln (b)} \biggr|_{b=1} = \frac{(2 \pi )^8 \bigl( \hat{\alpha}_{x}^{2} + \hat{\alpha}_{y}^{2} \bigr)}{2 \kappa_{xy}}~~.
\end{align}

It is also possible to formulate an integral corresponding to a quartic pure term. Specifically, let us imagine that the Gaussian part $\mathcal{L}_G$ in eq.~\eqref{partitionfunccc} gets modified so that it now includes the term: 
\begin{align} \label{eoiewo1111wiuwi}
\begin{split}
 \mathcal{L}^{\prime }_G &   =  \mathcal{L}_G + \frac{1}{2} \left[ \upsilon_x   (   \hat{\Delta}^2_x   h )^2 + \upsilon_y  (    \hat{\Delta}^2_y   h )^2   \right] ~~,
\end{split}
\end{align}
Using eq.~\eqref{eq:upsilonnew}, one finds that the generation of $\upsilon_x (b)$ is given by
\begin{align} \label{eq:upsilon}  
\begin{split}
      \upsilon_x (b) =   &  2 ( 2 \pi )^2  a_{x}^{2} \alpha_{x}^{2} (b) 
 \Omega_a \sum_i \, x_i^2 \Bigl( e^{g^{+}_{(xx)} (\mathbf{x}_i)} -1 \Bigr)  ~~ .
\end{split}
\end{align}
Following the same steps as were performed for $ \delta \kappa_{xy} (b)$, one finds that the integral over $p$ localizes the spatial sum at $x_i =0$, so that one obtains $  \upsilon_x (b) =0$ and the same for $\upsilon_y(b)$. A similar argument will also cause the vanishing of any integral generating non-fractonic coefficients at higher order in derivatives. We thus find agreement with the statement in Ref.~\cite{moessneryizhi} that dipole interactions only take place on the plane orthogonal to the dipole direction, thereby preventing screening effects that would break fractonic symmetry.

\section{RG flow diagram}
Let us return to the RG flow of the dipole fugacities. For simplicity, let us consider the isotropic case when $\alpha_x = \alpha_y$. This is a consistent truncation to the order of our RG calculations since the beta function of the difference $\alpha_x - \alpha_y$ is proportional to itself and so if it is set to 0 initially, then it does not subsequently get generated. Let us collectively denote these fugacities as
\begin{equation}
    \alpha_x = \alpha_y \equiv \alpha_d~~,
\end{equation}
where the subscript $d$ denotes ``dipole.''

\begin{figure}[t]
    \centering
    \includegraphics[width=0.4\textwidth]{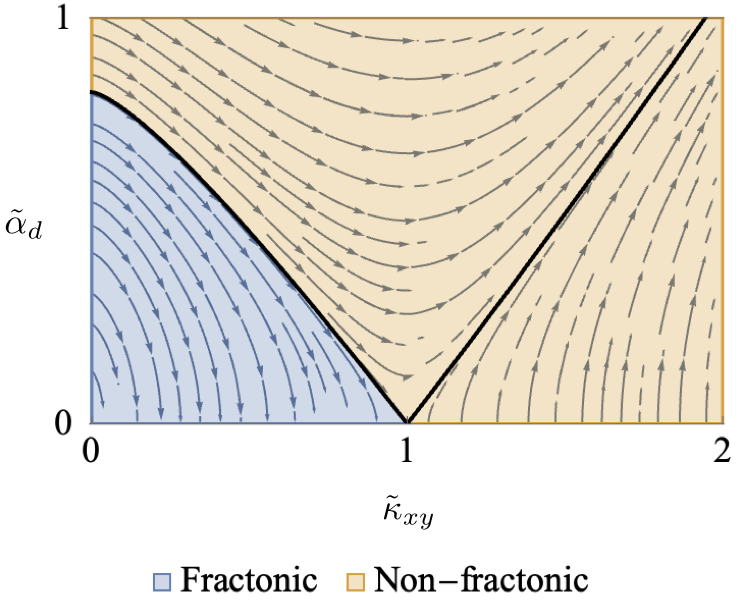}
    \caption{RG flow diagram for the rescaled coefficient $\tilde{\kappa}_{xy}$ and  fugacity $\tilde{\alpha}_{d}$.}
    \label{fig:rgflow}
\end{figure}

It is a simple matter to trace through the computation of $\alpha_x (b)$ to see what difference is made by splitting $\kappa$ up into $\kappa_{xy}$ and $\kappa_z$. The result for $\hat{\alpha}_d$ reads
\begin{equation}
    \frac{\partial \hat{\alpha}_d (b)}{\partial \ln (b)} \biggr|_{b=1} = \biggl( 2 - \frac{\pi^2}{2 \sqrt{\kappa_z \kappa_{xy}}} \biggr) \hat{\alpha}_d~~.
\end{equation}
If we define the rescaled variables
\begin{align}
    \tilde{\kappa}_{xy} &= \frac{16 \kappa_z \kappa_{xy}}{\pi^4}~~, &%
    \tilde{\alpha}_d &= 64 \kappa_z \hat{\alpha}_d~~,
\end{align}
then, the RG flow equations simplify to
\begin{subequations}
\begin{align}
    \frac{\partial \tilde{\kappa}_{xy} (b)}{\partial \ln (b)} \biggr|_{b=1} &= \frac{\tilde{\alpha}_{d}^{2}}{\tilde{\kappa}_{xy}}~~, \label{eq:kappatilderg} \\
    \frac{\partial \tilde{\alpha}_{d} (b)}{\partial \ln (b)} \biggr|_{b=1} &= 2 \tilde{\alpha}_{d} \biggl( 1 - \frac{1}{\sqrt{\tilde{\kappa}_{xy}}} \biggr)~~. \label{eq:alphatilderg}
\end{align}
\end{subequations}
%
The RG flow diagram is plotted in Fig. \ref{fig:rgflow} and the fractonic and non-fractonic regions are shaded accordingly. The critical trajectory, which flows to the fixed point from the left and away from the fixed point on the right, and is drawn as a thick black line in the plot, is determined by dividing eq. \eqref{eq:alphatilderg} by \eqref{eq:kappatilderg} to get $\frac{\partial \tilde{\alpha}_d}{\partial \tilde{\kappa}_{xy}}$ and solving the resulting differential equation for $\tilde{\alpha}_d$ as a function of $\tilde{\kappa}_{xy}$. This gives a generic trajectory parametrized by an arbitrary integration constant. Then, simply set the constant such that the trajectory passes through the fixed point. The result is
\begin{align}
    \tilde{\alpha}_{d}^{{\rm crit}} = \sqrt{\frac{2}{3} \bigl( 1 + 2 \sqrt{\tilde{\kappa}_{xy}} + 3 \tilde{\kappa}_{xy} \bigr)} \, \bigl| 1 - \sqrt{\tilde{\kappa}_{xy}} \bigr|~~.
\end{align}
Of course, we can only trust this perturbative calculation as long as $\tilde{\alpha}_d$ is small, which, for this critical trajectory, means that $\tilde{\kappa}_{xy}$ is close to 1. Around the critical point, the critical trajectory simplifies significantly to
\begin{align}
    \tilde{\alpha}_{d}^{{\rm crit}} \approx \bigl| \tilde{\kappa}_{xy} - 1 \bigr|~~.
\end{align}
The critical trajectory that flows towards the fixed point from the left and separates the fractonic and non-fractonic regions is also called a separatrix \cite{herbut_2007,fradkin2021quantum}.


\section{Conclusion} 

In this work, we looked at a fractonic BKT-transition in three dimensions that was first mentioned in Ref.~\cite{moessneryizhi}, where a fractonic plaquette-dimer liquid with algebraic correlations melts into a disordered phase without fractonic properties. With a momentum shell RG scheme we tailored for UV/IR mixing, we showed that one can derive the critical properties of this model. This is done through momentum shell RG by considering the renormalization of the cosine terms representing to defects in the dual theory. We also considered screening effects which happen at second order in fugacity. Specifically, we considered the possibility that the simple cosine operator, which represents monopole defects in the dual theory, act as a dangerously irrelevant operator that destroys the fractonic dispersion, thereby removing the UV/IR mixing nature of the model in the IR. We also considered screening effects related to terms quartic in derivatives, induced by dipole defects. Also at the quartic level, we find all screening effects generating non-fractonic couplings vanish, consistent with the statement in Ref.~\cite{moessneryizhi} that dipole interactions only exist transverse to the dipole direction. Moreover, we find that the coefficient of the fractonic term quartic in derivatives does get renormalized. We therefore were able to formulate an RG flow diagram which contains a low-temperature fractonic regime where one flows towards a fractonic phase. At higher temperatures, the fractonic phase gets destroyed through dipole proliferation, analogous to the ordinary BKT-transition. The RG scheme introduced in this work is an extension of Ref.~\cite{ethanlake}, which focused on computing the anomalous dimension of broad set of operators in case one has a Gaussian term with fractonic symmetry. In this work, the effect of defect screening was also considered, providing a more detailed understanding of the fractonic BKT transition of a dimer-plaquette model. This work opens the door to further exploring, from a renormalization group perspective, other models with the property that short wave length modes are part of the low-energy theory. \\

\emph{Acknowledgments.--}
We acknowledge discussions with Renato Dantas, Roderich Moessner, Andr\'{e}s Schlief Raether, Vijay Shenoy, and Yizhi You. We especially thank Ethan Lake for explaining  his work to us. KTG has received funding from the European Union’s Horizon 2020 research and innovation programme under the Marie Sk\l odowska-Curie grant agreement No 101024967. RL was supported, in part, by the cluster of excellence ct.qmat (EXC 2147, project-id 390858490). PS acknowledges the support of the Narodowe Centrum Nauki (NCN) Sonata Bis grant 2019/34/E/ST3/00405 and the Nederlandse Organisatie voor Wetenschappelijk Onderzoek (NWO) Klein grant via NWA route 2.

\onecolumngrid
\appendix

\section{Low-energy effective action}
   \label{partone}
In the RG computations performed in this appendix, we will closely follow Ref.~\cite{herbut_2007} (see also Refs.~\cite{Herbut_2003,seradjeh}), where a similar momentum shell RG computations is performed but for the standard BKT transition. A key difference in this appendix however, is that this appendix does not work in the continuum limit, as the continuum limit is not as trivial for fractonic models and it is best to consider the continuum limit at a later stage when the integral has been rewritten in such a way that the lattice constant no longer plays a role except as part of a Riemann sum.

We start with the partition function of our model expanded near the fractonic free-field fixed point:
\begin{align} \label{partitionfuncccapp}
    \mathcal{Z} &=  \int^{\infty}_{-\infty} \prod_i  d h_i  \, e^{- S }, &%
    S &= \Omega_a  (  K + V) ,
\end{align}
where $S$ is the static action and $K$ and $V$ are the kinetic and potential energies,
\begin{align}
    K &=   \sum_i \,  \frac{\kappa_{xy}}{2} \bigl( \Delta_x \Delta_y h_i \bigr)^2 + 
 \frac{\kappa_{z}}{2}\bigl( \Delta_z h_i \bigr)^2 , &%
    V &= 2 \sum_i \sum_I \alpha_I \cos ( 2 \pi f_{I i} ) ~~ . 
\end{align}
First, let us expand out the contribution of potential to the partition function:
\begin{align}
    \mathcal{Z} &=  \int^{\infty}_{-\infty} \prod_i  d h_i  \, e^{- \Omega_a K} \sum_{n=0}^{\infty} \frac{  \Omega_a^n  V^n}{n!}~~.
\end{align}
To integrate out the momentum shells, as usual, we divide the field $h$ into high- and low-energy parts:
   \begin{align}
     h   = h^+  + h^-  ~~ .  \label{divideapp}
   \end{align}

   The modes $h^+$ have energies within some energy shell between energies $\Lambda / b$ and $\Lambda$, where $b$ is a number slightly greater than 1. We integrate these modes out leaving behind just the low-energy modes $h^-$. The remaining momenta have to then be rescaled back up from $\Lambda / b$ to $\Lambda$. The latter step similarly scales the couplings according to their scaling dimension. As we have argued in the main text, the scaling that is appropriate near the fractonic fixed point is a type of dilatation towards the $x$ and $y$ axes, as opposed to towards the origin.
   
   Since $h^-$ and $h^{+}$ have no overlap there are no cross terms between them arising from the free kinetic part of the action. However, we must still expand the cosine terms:
\begin{align}
    \cos ( 2 \pi f_I ) &= \cos \bigl[ 2 \pi \bigl( f_{I}^{+} + f_{I}^{-} \bigr) \bigr] = \cos ( 2 \pi f_{I}^{+} ) \, \cos ( 2 \pi f_{I}^{-} ) - \sin ( 2 \pi f_{I}^{+} ) \, \sin ( 2 \pi f_{I}^{-} )~~.
\end{align}
The partition function now reads
\begin{align}
    \mathcal{Z} &= \int \mathcal{D} h^- \, e^{- 
 \Omega_a K_L} \int \mathcal{D} h^{+} \, e^{- \Omega_a K^{+}} \biggl\{ 1 + 2  \Omega_a \sum_{i,I} \alpha_{I} \, \bigl[ \cos ( 2 \pi f_{I}^{+} ) \, \cos ( 2 \pi f_{I}^{-} ) - \sin ( 2 \pi f_{I}^{+} ) \, \sin ( 2 \pi f_{I}^{-} ) \bigr] (\mathbf{x}_i)  \notag \\
    &\quad + 2  \Omega_a^2  \sum_{i, j,I,J } \alpha_I \alpha_J  \, \bigl[ \cos ( 2 \pi f_{I}^{+} ) \, \cos ( 2 \pi f_{I}^{-} ) - \sin ( 2 \pi f_{I}^{+} ) \, \sin ( 2 \pi f_{I}^{-} ) \bigr] (\mathbf{x}_i) \notag \\
    &\hspace{3.75cm} \times \bigl[ \cos ( 2 \pi f_{J}^{+} ) \, \cos ( 2 \pi f_{J}^{-} ) - \sin ( 2 \pi f_{J}^{+} ) \, \sin ( 2 \pi f_{J}^{-} ) \bigr] (\mathbf{x}_j) + O (V^3) \biggr\}~~, 
\end{align}
where we defined $\int^{\infty}_{-\infty} \prod_i  d h_i = \int \mathcal{D} h$. 
Integrating out $h^{+}$ amounts to taking the expectation value with respect to $K^{+}$, which we denote by
\begin{equation}
    \langle \mathcal{O} \rangle^{+} = \frac{\int \mathcal{D} h^{+} \, e^{-K^{+}} \, \mathcal{O}}{\int \mathcal{D} h^{+} \, e^{-K^{+}}}~~,
\end{equation}
for an arbitrary functional $\mathcal{O}$ of $h^{+}$.

Since there is no tadpole for $h^{+}$ (no term linear in $h^{+}$), the expectation value of any function which is odd in $h^{+}$ vanishes. In other words,
\begin{align}
       \bigl\langle   \sin \bigl[ 2 \pi  f_{I}^{+} (\mathbf{x}_i) \bigr] \bigr\rangle^+   =
           \bigl\langle   \sin \bigl[ 2 \pi  f_{I}^{+} (\mathbf{x}_i) \bigr]  \cos \bigl[  2 \pi  f_{J}^{+} ( \mathbf{x}_j) \bigr]    \bigr\rangle^+   =0  ~~ . 
\end{align}
Therefore, the only terms which survive in the partition function are
\begin{align}
    \mathcal{Z}^{-} &= \int \mathcal{D} h^- \, e^{- 
 \Omega_a K^{-}} \biggl\{ 1 + 2  \Omega_a \sum_{i,I} \alpha_I  \, \cos \bigl( 2 \pi f_{I}^{-} \bigr) \bigl\langle \cos \bigl( 2 \pi f_{I}^{+} \bigr) \bigr\rangle^+ \notag \\
    &\quad + 2  \Omega_a^2  \sum_{i,j,I,J} \alpha_I \alpha_J   \, \cos \bigl[ 2 \pi f_{I}^{-} (\mathbf{x}_i) \bigr] \cos \bigl[ 2 \pi f_{J}^{-} (\mathbf{x}_j) \bigr] \bigl\langle \cos \bigl[ 2 \pi f_{I}^{+} (\mathbf{x}_i) \bigr] \cos \bigl[ 2 \pi f_{J}^{+} (\mathbf{x}_j) \bigr] \bigr\rangle^+ \notag \\
    &\quad + 2  \Omega_a^2  \sum_{i,j,I,J} \alpha_I \alpha_J \, \sin \bigl[ 2 \pi f_{I}^{-} (\mathbf{x}_i) \bigr] \sin \bigl[ 2 \pi f_{J}^{-} (\mathbf{x}_j) \bigr] \bigl\langle \sin \bigl[ 2 \pi f_{I}^{+} (\mathbf{x}_i) \bigr] \sin \bigl[ 2 \pi f_{J}^{+} (\mathbf{x}_j) \bigr] \bigr\rangle^+ + O (V^3) \biggr\}~~.
\end{align}
The expectation values of these trigonometric functions are related to the basic 2-point function
\begin{equation}
    g^{+}_{(IJ)} (\mathbf{x}_i) = (2 \pi)^2 \bigl\langle f_{I}^{+} (\mathbf{x}_i) \, f_{J}^{+} (0) \bigr\rangle^+~~,
\end{equation}
via Wick's theorem:
\begin{subequations}
\begin{align}
    \bigl\langle \cos \bigl[ 2 \pi f_{I}^{+} (\mathbf{x}_i) \bigr] \bigr\rangle^+ &= e^{- \frac{1}{2} g^{+}_{(II)} (0)}~~, \\
    \bigl\langle \cos \bigl[ 2 \pi f_{I}^{+} (\mathbf{x}_i) \bigr] \cos \bigl[ 2 \pi f_{J}^{+} (\mathbf{x}_j) \bigr] \bigr\rangle^+ &= e^{- \frac{1}{2} \bigl[ g^{+}_{(II)} (0) + g^{+}_{(JJ)} (0) \bigr]} \cosh \bigl[ g^{+}_{(IJ)} (\mathbf{x}_i-\mathbf{x}_j) \bigr]~~, \\
    \bigl\langle \sin \bigl[ 2 \pi f_{I}^{+} (\mathbf{x}_i) \bigr] \sin \bigl[ 2 \pi f_{J}^{+} (\mathbf{x}_j) \bigr] \bigr\rangle^+ &= e^{- \frac{1}{2} \bigl[ g^{+}_{(II)} (0) + g^{+}_{(JJ)} (0) \bigr]} \sinh \bigl[ g^{+}_{(IJ)} (\mathbf{x}_i-\mathbf{x}_j) \bigr]~~.
\end{align}
\end{subequations}
Therefore, the low-energy partition function reads
\begin{align}
    \mathcal{Z}^{-} &= \int \mathcal{D} h^- \, e^{- \Omega_a K^{-}} \biggl\{ 1 + 2  \Omega_a \sum_{i,I} \alpha_I e^{- \frac{1}{2} g^{+}_{(II)} (0)}  \, \cos \bigl( 2 \pi f_{I}^{-} \bigr) \notag \\
    &\quad + 2    \Omega_a^2  \sum_{i,j ,I,J} \alpha_I \alpha_J e^{- g^{+}_{(IJ)} (0)}   \, \cos \bigl[ 2 \pi f_{I}^{-} (\mathbf{x}_i) \bigr] \cos \bigl[ 2 \pi f_{J}^{-} (\mathbf{x}_j) \bigr] \cosh \bigl[ g^{+}_{(IJ)} (\mathbf{x}_i - \mathbf{x}_j ) \bigr] \notag \\
    &\quad + 2   \Omega_a^2  \sum_{i,j,I,J} \alpha_I \alpha_J e^{- g^{+}_{(IJ)} (0)}  \, \sin \bigl[ 2 \pi f_{I}^{-} (\mathbf{x}_i) \bigr] \sin \bigl[ 2 \pi f_{J}^{-} (\mathbf{x}_j) \bigr] \sinh \bigl[ g^{+}_{(IJ)} (\mathbf{x}_i - \mathbf{x}_j ) \bigr] + O (V^3) \biggr\} ~~.
\end{align}
We can re-exponentiate this to define a low-energy potential term,
\begin{align}
    V^{-} &= 2  \Omega_a \sum_i \sum_I \alpha_I e^{- \frac{1}{2} g^{+}_{(II)} (0)} \cos \bigl( 2 \pi f_{I}^{-} \bigr) \notag \\
    &\quad + 2 \Omega_a^2   \sum_{i,j,I,J} \alpha_I \alpha_J e^{- \frac{1}{2} \bigl[ g^{+}_{(II)} (0) + g^{+}_{(JJ)} (0) \bigr]} \biggl\{ \cos \bigl[ 2 \pi f_{I}^{-} (\mathbf{x}_i) \bigr] \cos \bigl[ 2 \pi f_{J}^{-} (\mathbf{x}_j) \bigr] \cosh \bigl[ g^{+}_{(IJ)} (\mathbf{x}_i-\mathbf{x}_j) \bigr] \notag \\
    &\hspace{1cm} + \sin \bigl[ 2 \pi f_{I}^{-} (\mathbf{x}_i) \bigr] \sin \bigl[ 2 \pi f_{J}^{-} (\mathbf{x}_j) \bigr] \sinh \bigl[ g^{+}_{(IJ)} (\mathbf{x}_i-\mathbf{x}_j) \bigr] - \cos \bigl[ 2 \pi f_{I}^{-} (\mathbf{x}_i) \bigr] \, \cos \bigl[ 2 \pi f_{J}^{-} (\mathbf{x}_j) \bigr] \biggr\} + \cdots~~,
\end{align}
where $\cdots$ are higher-order terms involving more factors of $\alpha_I$. We can massage this into the form
\begin{align}
    V^{-} &= 2  \Omega_a \sum_{i , I} \alpha_I e^{- \frac{1}{2} g^{+}_{(II)} (0)} \cos \bigl( 2 \pi f_{I}^{-} \bigr) \notag \\
    &\quad + 2  \Omega_a^2  \sum_{i,j,I,J} \alpha_I \alpha_J e^{- \frac{1}{2} \bigl[ g^{+}_{(II)} (0) + g^{+}_{(JJ)} (0) \bigr]} \biggl\{ \cos \bigl[ 2 \pi \bigl( f_{I}^{-} (\mathbf{x}_i) + f_{J}^{-} (\mathbf{x}_j) \bigr) \bigr] \Bigl( e^{- g^{+}_{(IJ)} (\mathbf{x}_i- \mathbf{x}_j)} -1 \Bigr) \notag \\
    &\hspace{7cm} + \cos \bigl[ 2 \pi \bigl( f_{I}^{-} (\mathbf{x}_i) - f_{J}^{-} (\mathbf{x}_j) \bigr) \bigr] \Bigl( e^{g^{+}_{(IJ)} (\mathbf{x}_i- \mathbf{x}_j)} -1 \Bigr) \biggr\} ~~.
\end{align}
Finally, we rescale momenta by a factor of $b$ in order to bring the cut-off back up from $\Lambda / b$ to $\Lambda$ and be able to compare the renormalized action to the original one. Each factor of $\alpha_I$ gets multiplied by $b^2$ since they have dimension 2, as argued in the main text. Let $V(b)$ be the renormalized potential as a function of $b$:
\begin{align}
    V (b) &= 2   \Omega_a \sum_i \sum_I \alpha_I \, b^2 e^{- \frac{1}{2} g^{+}_{(II)} (0)} \cos \bigl( 2 \pi f_{I}^{-} \bigr) \notag \\
    &\quad + 2   \Omega_a^2  \sum_{i ,j,I,J} \alpha_I \alpha_J \, b^4 e^{- \frac{1}{2} \bigl[ g^{+}_{(II)} (0) + g^{+}_{(JJ)} (0) \bigr]} \biggl\{ \cos \bigl[ 2 \pi \bigl( f_{I}^{-} (\mathbf{x}_i) + f_{J}^{-} (\mathbf{x}_j) \bigr) \bigr] \Bigl( e^{- g^{+}_{(IJ)} (\mathbf{x}_i- \mathbf{x}_j)} -1 \Bigr) \notag \\
    &\hspace{7.4cm} + \cos \bigl[ 2 \pi \bigl( f_{I}^{-} (\mathbf{x}_i) - f_{J}^{-} (\mathbf{x}_j) \bigr) \bigr] \Bigl( e^{g^{+}_{(IJ)} (\mathbf{x}_i- \mathbf{x}_j)} -1 \Bigr) ~~.
\end{align}
The renormalization of the original cosine terms in the potential are retrieved from the first line above and we will focus on these terms in the following sections. The renormalized fugacities read
\begin{equation} \label{eq:alpharenorm}
     \alpha_I (b)= \alpha_I b^2 e^{- \frac{1}{2} g^{+}_{(II)} (0)}~~.
\end{equation}
This is an implicit equation that requires the evaluation of $\alpha_I$ at some scale. As usual, what we are interested in is the beta function, which is the logarithmic derivative with respect to $b$.\footnote{Our definition of the beta function is negative of the standard one because by some historical quirk the usual definition of the beta function flows towards the UV.}

Importantly, however, additional terms are generated in the subsequent lines of $V(b)$. This is not surprising, this happens also in the usual BKT analysis in two dimensions. The terms in the second line introduce so-called ``higher harmonics'' (e.g., when $I=J=0$ these are vortices of vorticity number greater than 1). These terms are less relevant and will be ignored for the same reason as in the standard case \cite{herbut_2007}. The terms in the expansion of the final line around $\mathbf{x}_i = \mathbf{x}_j$, at least for $I=J$, constitute corrections to kinetic terms, which we will consider. 

Let us write out this term in general. First, shift $\mathbf{x}_i$ to $\mathbf{x} - \mathbf{x}'$ and expand in $\mathbf{x}'$:
\begin{align}
    V_I (b) &\equiv 2  \Omega_a \sum_{i , j,I} \alpha_{I}^{2} b^4 e^{-g^{+}_{(II)} (0)} \cos \bigl[ 2 \pi \bigl( f_{I}^{-} (x) - f_{I}^{-} (\mathbf{x}_j) \bigr] \Bigl( e^{g^{+}_{(II)} (x-\mathbf{x}_j)} - 1 \Bigr) \notag \\
    &\approx 2  \Omega_a^2  \sum_{i , j,I} \alpha_{I}^{2} (b) \cos \bigl[ 2 \pi \mathbf{x}_i \cdot \hat{\mathbf{\Delta}} f_{I}^{-} (x) \bigr] \Bigl( e^{g^{+}_{(II)} (\mathbf{x}_j)} - 1 \Bigr) \notag \\
    &\approx 2  \Omega_a^2  \sum_{i , j,I} \alpha_{I}^{2} (b) \biggl( 1 - \frac{1}{2} \bigl[ 2 \pi \mathbf{x} \cdot \hat{ \mathbf{\Delta}}  f_{I}^{-} (x) \bigr]^2 \biggr) \Bigl( e^{g^{+}_{(II)} (\mathbf{x}_j)} - 1 \Bigr)~~.
\end{align}
Let $V_{\mu I}$ be the contribution to the kinetic operator $- \frac{1}{2} \bigl[ \hat{ \Delta}_{\mu} f_{I}^{+} (\mathbf{x}_i) \bigr]^2$. Then, for example,
\begin{align}
    V_{x0} &= - \frac{1}{2} \, 2 (2 \pi )^2 \alpha_{0}^{2} (b)  \Omega^2_a \sum_j  \,  x_j^2 \Bigl( e^{g^{+}_{(00)} (\mathbf{x}_j)} -1 \Bigr) \sum_i ( \hat{\Delta}_x h^- (\mathbf{x}_i ) )^2~~.
\end{align}
By $x_j \leftrightarrow y_j $ symmetry, we can replace $x_j^2$ with $\frac{ x_j^2 + y_j^2}{2}$. In this way, it is clear that the coefficient of $( \hat{\Delta}_y h^- )^2$ in $V_{y0}$ is exactly the same as the coefficient of $( \hat{\Delta}_x h^- )^2$ in $V_{x0}$. Therefore, the contribution to a putative $\gamma \bigl[ ( \hat{\Delta}_x h )^2 + ( \hat{\Delta}_y h )^2 \bigr]$ term in the Lagrangian from $\alpha_0$ is given by
\begin{align} \label{eq:gammaeq}
     \gamma (b)  &= ( 2 \pi )^2 \alpha_{0}^{2} (b)  \Omega_a \sum_i \, (x_i^2 + y_i^2 ) \Bigl( e^{g^{+}_{(00)} (\mathbf{x}_i)} -1 \Bigr)~~.
\end{align}
Similarly, consider
\begin{align}
    V_{yx} &= - \frac{1}{2} \, 2 ( 2 \pi )^2 a_{x}^{2} \,  \alpha_{x}^{2} (b)  \Omega_a^2  \sum_j \, y^2 \Bigl( e^{g^{+}_{(xx)} (\mathbf{x}_j)} - 1 \Bigr) \sum_i \, ( \hat{\Delta}_x  \hat{\Delta}_y h^-  (\mathbf{x}_i)  )^2~~.
\end{align}
Caution: we do \emph{not} have $x_i \leftrightarrow y_i$ symmetry here! However, we do have another term:
\begin{align}
    V_{xy} &= - \frac{1}{2} \, 2 ( 2 \pi )^2 a_{y}^{2} \alpha_{y}^{2} (b)  \Omega_a^2  \sum_j \, x_j^2 \Bigl( e^{g^{+}_{(yy)} (\mathbf{x}_j)} - 1 \Bigr) \sum_i \, ( \hat{\Delta}_x \hat{\Delta}_y h^- (\mathbf{x}_i) )^2~~.
\end{align}
Then, the flow of $\kappa_{xy}$ is given by
\begin{align} \label{eq:kappaxyeq}
\begin{split}
    \delta \kappa_{xy} (b) \equiv \kappa_{xy} (b) - \kappa_{xy}  & =    2 ( 2 \pi )^2  a_{x}^{2} \alpha_{x}^{2} (b)  \Omega_a \sum_i \, y_i^2 \Bigl( e^{g^{+}_{(xx)} (\mathbf{x}_i)} -1 \Bigr) + x \leftrightarrow y  ~~.  
\end{split}
  \end{align}
Lastly, we consider $V_{xx}$ or $V_{yy}$:
\begin{align}
    V_{xx}  = - \frac{1}{2} \, 2 ( 2 \pi )^2 a_{x}^{2} \alpha_{x}^{2} (b)  \Omega_a^2  \sum_j \, x_j^2 \Bigl( e^{g^{+}_{(xx)} (\mathbf{x}_j)} - 1 \Bigr) \sum_i \, ( \hat{\Delta}^2_x h^- (\mathbf{x}_i) )^2  ~~ , \\ 
     V_{yy}  = - \frac{1}{2} \, 2 ( 2 \pi )^2 a_{y}^{2} \alpha_{y}^{2} (b)  \Omega_a^2  \sum_j \, y_j^2 \Bigl( e^{g^{+}_{(yy)} (\mathbf{x}_j)} - 1 \Bigr) \sum_i \, (  \hat{\Delta}^2_y h^- (\mathbf{x}_i) )^2 ~~ . 
\end{align}
From $  V_{xx}$, we can obtain the coefficient $\upsilon_x (b)$ of the non-fractonic operator:
\begin{align} \label{eq:upsilonnew}  
\begin{split}
      \upsilon_x (b) =   &  2 ( 2 \pi )^2  a_{x}^{2} \alpha_{x}^{2} (b) 
 \Omega_a \sum_i \, x_i^2 \Bigl( e^{g^{+}_{(xx)} (\mathbf{x}_i)} -1 \Bigr)  ~~ .
\end{split}
\end{align}


\section{Taking note of the discreteness of the gradients set by the lattice scale}
\label{sec:discrete}
In this appendix we revisit the simplification of eq.~\eqref{eq:dispersionsimple}. As we show, this simplification is quantitatively invalid, because eq.~\eqref{eq:simplification} is not valid for the fractonic momentum shells that are integrated over in this work. If we do not perform this simplification, the momentum shell is given by
\begin{align} \label{eiei2-0922-=-}
    \Lambda^2 =  \frac{ 16 \sin^2 ( a_x p /2 ) \sin^2 ( a_y  q /2  )}{ a^2_x a^2_y }  + k^2 ~~.  
\end{align}
Note that in eq.~\eqref{eiei2-0922-=-} the $k$-momentum is still expanded, This is because for this direction there is no concern for the expansion being inaccurate, as the phenomenon of UV/IR mixing is restricted to the $pq$-plane in this model. Then, the simple Gaussian correlator in momentum space, which one needs for the renormalization group analysis, is modified by
\begin{align}
  \langle  f_0 (\mathbf{p}) f_0 (0 )  \rangle =   \frac{1 }{\kappa (p^2 q^2 + k^2 )} \rightarrow   \frac{1}{\kappa} \frac{1}{  \frac{16 \sin^2 ( a_x p /2 ) \sin^2 (a_y  q/2  )}{ a^2_x a^2_y }  + k^2 } ~~.
\end{align}
Similarly
\begin{subequations}
\begin{align}
  \langle  f_x (\mathbf{p}) f_x (0 )  \rangle  & =   \frac{ a_x^2  p^2 }{\kappa (p^2 q^2 + k^2 )} \rightarrow   \frac{1 }{\kappa } \frac{4 \sin^2 ( a_x p/2 )}{  \frac{16 \sin^2 ( a_x p /2 ) \sin^2 (a_y  q/2  )}{ a^2_x a^2_y }  + k^2 } ~~ , \label{3e2209209202} \\ 
    \langle  f_y (\mathbf{p}) f_y (0 )  \rangle  & =   \frac{ a_y^2  q^2  }{\kappa (p^2 q^2 + k^2 )} \rightarrow   \frac{1}{\kappa } \frac{4 \sin^2 ( a_y  q/2  )}{  \frac{16 \sin^2 ( a_x p /2 ) \sin^2 (a_y  q/2  )}{ a^2_x a^2_y }  + k^2 } ~~ . 
\end{align}                
\end{subequations}
We are most interested in finding the quantitative effect of the discretization on the critical value for the coefficient $\kappa$ above which a transition takes place. For this we use Eq.~\eqref{3e2209209202} to find
\begin{align} \label{feieoi2010-10-}
      g_{(xx)}^{+} ( 0 )  &=  \frac{(2\pi)^2   }{\kappa}  \int_{\shell} \frac{d^3 \mathbf{p}}{(2 \pi)^3 }  \frac{4 \sin^2 ( a_x p/2 )}{  \frac{16 \sin^2 ( a_x p /2 ) \sin^2 (a_y  q/2  )}{ a^2_x a^2_y }  + k^2 }   ~~ . 
\end{align}
Again, we pass to dimensionless variables as in Eq.~\eqref{eq:undim} so that Eq.~\eqref{feieoi2010-10-} turns into
\begin{align} \label{feieoi2010-10-3}
      g_{(xx)}^{+} ( 0 )  &=  \frac{ \pi^3     }{ 2 \kappa}  \int_{\shell}   d^3  \mathbf{ \tilde p}  \frac{4 \sin^2 ( \pi  \tilde p/2 )}{ 16 \sin^2 ( \pi \tilde  p /2 ) \sin^2 ( \pi \tilde  q/2  )  + \pi^4 \tilde k^2 }   ~~ . 
\end{align}
It is difficult to evaluate Eq.~\eqref{feieoi2010-10-3} analytically because we have to take note of the momentum shell given in Eq.~\eqref{eiei2-0922-=-} and we therefore simplify in two ways. Firstly, as in eq.~\eqref{eq:integralsimplification1}, we consider a momentum shell along the $pq$-plane, so that we can immediately take the continuum limit for the $z$-direction and integrate out $ k $. Eq.~\eqref{feieoi2010-10-3} then reduces to
\begin{align} \label{feieoi2010-1398398}
      g_{(xx)}^{+} ( 0 )  &=  \frac{2  \pi     }{ \kappa}   \int_{\shell}    d \tilde p  d \tilde q   \frac{ | \sin ( \pi  \tilde p/2 ) | }{   \tilde  q    }   ~~ . 
\end{align}
Secondly, like in fig.~\ref{fig:esurfaces1111}, we simplify the shell by on the $pq$-plane by turning it into straight lines tangent to the $\tilde p$-direction. We then find
\begin{align}
\begin{split}
    \label{298289982}
      g_{(xx)}^{+} ( 0 )   & \approx    \frac{ 2   \pi    }{    \kappa}   \int_{-1}^{1 }  d \tilde p      | \sin ( \pi  \tilde p/2 ) | \ln(b) =     \frac{8  }{\kappa } \ln(b)  ~~.   
\end{split}
   \end{align}
From Eqs.~\eqref{298289982} it follows that the flow of $\alpha_x$ is given by 
 \begin{align} \label{f3h0301}
      \frac{\partial \alpha_{x} (b)}{\partial \ln (b)} \biggr|_{b=1} =  2 \biggl(1 -  \frac{2 }{  \kappa}  \biggr) \alpha_x  ~~ .
  \end{align}
  So one learns that the critical point for this discretized computation is given by
  \begin{align}
      \kappa^{(L)}_c  =  2 ~~. 
  \end{align}

\section{Renormalization of the simple cosine term}
\label{sec:cos}

For the simple cosine term, which is when $I=0$ and $f_0 = h$, we have
\begin{align}
    g^{+}_{(00)} (0) = \lim_{x' \rightarrow x} ( 2 \pi )^2 \bigl\langle h(x) h(x') \bigr\rangle = ( 2 \pi )^2 \int \frac{d^3 \mathbf{p}}{( 2 \pi )^3} \frac{1}{\kappa \bigl[ (pq)^2 + k^2 \bigr]}~~,
\end{align}
In dimensionless variables, the expression of the integral actually remains the same, just with tildes on the variables:
\begin{align}
    g^{+}_{(00)} (0) = \frac{1}{2 \pi \kappa} \int \frac{d^3 \tilde{\mathbf{p}}}{( \tilde{p} \tilde{q} )^2 + \tilde{k}^2} = \frac{4}{\pi \kappa} \int_{>0} \frac{d \tilde{k} \, d \tilde{\ell}}{\tilde{k}^2 + \tilde{\ell}^2} \int_{\tilde{\ell}}^{1} \frac{d \tilde{p}}{\tilde{p}} = \frac{4}{\pi \kappa} \int_{>0} \frac{d \tilde{k} \, d \tilde{\ell}}{\tilde{k}^2 + \tilde{\ell}^2} \, \ln \frac{1}{\tilde{\ell}}~~.
\end{align}
Again, the subscript $>0$ on the integral means restrict to the positive octant or quadrant. In radial coordinates,
\begin{align}
    g^{+}_{(00)} (0) &= \frac{4}{\pi \kappa} \int_{\tilde{\Lambda} / b}^{\tilde{\Lambda}} \frac{d \tilde{\lambda}}{\tilde{\lambda}} \int_{0}^{\pi / 2} d \theta \, \ln \biggl( \frac{1}{\tilde{\lambda} \, \sin \theta} \biggr) \notag \\
    &= \frac{2}{\kappa} \int_{\tilde{\Lambda} / b}^{\tilde{\Lambda}} \frac{d \tilde{\lambda}}{\tilde{\lambda}} \, \ln \biggl( \frac{2}{\tilde{\lambda}} \biggr) \notag \\
    &= \frac{2}{\kappa} \biggl[ \ln \biggl( \frac{2}{\tilde{\Lambda}} \biggr) + \frac{1}{2} \ln (b) \biggr] \ln (b)~~.
\end{align}
Therefore, the renormalized fugacity is
\begin{align}
    \alpha_0 (b) &= \alpha_0 \, b^{2 - \frac{1}{\kappa} \bigl[ \ln \bigl( \frac{2}{\tilde{\Lambda}} \bigr) + \frac{1}{2} \ln (b) \bigr]}~~,
\end{align}
and the logarithmic derivative at $b=1$ evaluates to\footnote{Unlike in relativistic theories in which multilogs appear only at two and higher loops, here a multilog has appeared already at one loop. We treat this in precisely the same way as in nonrelativistic quantum critical systems at finite density \cite{Fitzpatrick:2014efa, Fitzpatrick:2014cfa}. Notably, the example of fermions near a Fermi surface interacting with a gapless boson studied therein also exhibits UV/IR mixing.}
\begin{align}
    \frac{d \alpha_0 (b)}{d \ln (b)} \biggr|_{b=1} = \biggl[ 2 - \frac{1}{\kappa} \ln \biggl( \frac{2}{\tilde{\Lambda}} \biggr) \biggr] \alpha_0~~.
\end{align}
The key here is that, since $\tilde{\Lambda} \ll 1$, it follows that $\alpha_0$ decays very quickly under the RG flow towards the infrared. Thus, around the fractonic fixed point, this operator is irrelevant. This operator does not destabilize the fractonic phase.

\bibliographystyle{utphys2}
\bibliography{RGbib}

\end{document}